\documentclass{iaus}
\usepackage{graphicx}

\title[Is the wind of the Oe star HD\,155806 magnetically confined?] 
{Is the wind of the Oe star HD\,155806 magnetically confined?}

\author[V. Petit, A. W. Fullerton, S. Bagnulo, G. A. Wade, MiMeS Collaboration]   
{V. Petit$^1$, A. W. Fullerton$^2$, S. Bagnulo$^3$, G. A. Wade$^4$ \and the MiMeS Collaboration}

\affiliation{$^1$D\'epartement de Physique, Universit\'e Laval, Qu\'ebec, Canada \break$^2$Space Telescope Science Institute, Baltimore, USA \break $^3$Armagh Observatory, Armagh, Northern Ireland\break$^4$Department Physics, Royal Military College of Canada, Kingston, Canada}

\pubyear{2009}
\volume{259}  
\pagerange{100--100}
\date{?? and in revised form ??}
\jname{Cosmic Magnetic Fields: From Planets, to Stars and Galaxies}
\editors{K.G. Strassmeier, A.G. Kosovichev \& J.E. Beckman, eds.}
\begin{document}

\maketitle

\begin{abstract}
Oe stars are a subset of the O-type stars that exhibit emission lines from a circumstellar disk. The recent detection of magnetic fields in some O-type stars suggests a possible explanation for the stability of disk-like structures around Oe stars. According to this hypothesis, the wind of the star is channeled by a dipolar magnetic field producing a disc in the magnetic equatorial plane.
As a test of this model, we have obtained spectropolarimetric observations of the hottest Galactic Oe star HD 155806. Here we discuss the results and implications of those observations.
\keywords{Stars: magnetic fields, stars: mass loss, stars: early-type, stars: emission-line, Be}
\end{abstract}

\begin{figure}
\centering
 \includegraphics[width=13cm]{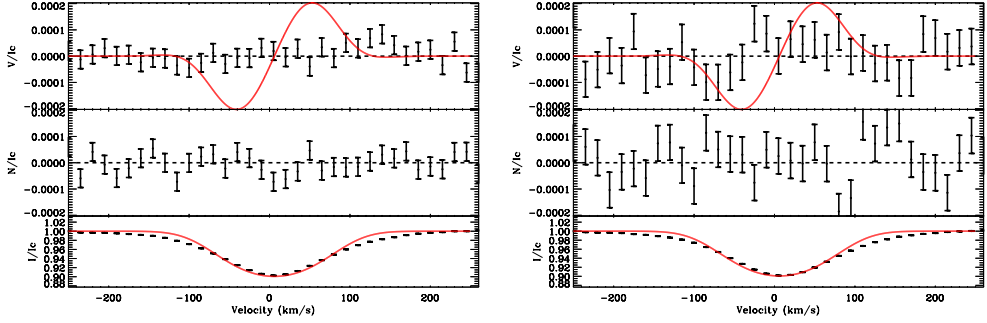}
  \caption{Least Squares Deconvolved profiles in June (left) and July (right).  The curves are the mean Stokes I profiles (bottom), the mean Stokes V profiles (top) and the N diagnostic null profiles (middle). The bold line represents the smallest-amplitude Stokes V profile corresponding to the longitudinal field of 115 G reported by Hubrig et al. (2007), which is a pure dipole with a polar strength of 380\,G.}\label{fig:fig1}
\end{figure}

Oe stars are a subset of the O-type stars that exhibit emission in spectra of their Balmer lines, indicating the presence of a circumstellar disk.
It is generally believed that the low incidence of the Oe stars (only 4 are known in the Milky Way) is caused by the increasingly powerful radiation-driven winds of the O-type stars, which inhibit the formation of long-lived circumstellar structures. However, all the \textit{bona fide} Oe stars exhibit typical wind profiles in their ultraviolet resonance lines, showing that these stars have winds similar to those of other O-type stars. The mystery is therefore not ``why are there so few Oe stars?'', but ``why are there any at all?''. 

The recent detection of magnetic fields in O-type stars suggests a possible explanation for the stability of disk-like structures around Oe stars. From MHD simulations, ud-Doula et al. (2008) have shown that under high magnetic confinement and high rotation, a ``magnetically confined wind'' can trap material in its magnetosphere. The circumstellar structure is composed of different material at different times, but its longevity is mediated by the magnetic field of the star. 
Consequently, this model neatly avoids problems associated with the disruption of the disks of Oe stars, and suggests that Oe stars might represent extreme cases of a more widespread phenomenon that depends on the balance between the strength of the stellar wind and the strength of the magnetic field.

HD\,155806 is the Galactic Oe star with the most powerful wind. We estimated that a surface dipolar magnetic field of at least 235 G is required to confine its wind. It exhibits a large, single-peaked H$\alpha$ emission profile that implies that the disk is not being viewed ``edge-on''. 
Encouragingly, the detection of a weak magnetic field ($115\pm37$\,G) in HD\,155806 with VLT/FORS1 has recently been reported by Hubrig et al. (2007), although this detection was not reproduced in further measurements (\cite{2008A&A...490..793H}). In addition, a re-analysis of the FORS1 observations (available in ESO archives) as a component of this present study led to less conclusive results. 
New high-resolution spectropolarimetric observations of HD\,155806 were obtained with ESPaDOnS at CFHT in June and July 2008. No magnetic Stokes V signature is detected in our observations (Figure 1). In order to extract the surface field characteristics constrained by the observed Stokes V profiles, we are in the process of comparing them with theoretical profiles derived from a large grid of dipolar magnetic field configurations that were calculated with the polarised LTE radiative transfer code Zeeman2 (\cite{1988ApJ...326..967L}). With the Bayesian method described by Petit et al. (2008), we will obtain an upper limit on the strength of the dipolar magnetic field at the stellar surface that is independent of the magnetic configuration. We shall also use the FORS1 measurements as additional constraints. 

Although detailed modeling still needs to be performed, we estimate that the upper limit on the surface magnetic dipole will be about 200\,G, according to our preliminary analysis. 
A stronger magnetic field would need to be in a configuration that produces a weak or null Stokes V signature. For dipoles, these rare configurations have well-defined geometries and correspond to the observer looking directly at the magnetic equator.
According to MHD simulations of an aligned dipole rotator (\cite{2008MNRAS.385...97U}), an accumulation at the equator of the magnetosphere will occur when both the confinement parameter and the rotation rate are large. Rigid-field hydrodynamics modelling of a tilted dipole (\cite{2007MNRAS.382..139T}) predicts that the disk will have an average inclination that lies somewhere between the rotation and magnetic equatorial planes, with plasma concentrated at the intersection between them. If the disk of HD\,155806 is produced by a large magnetic field hidden in a Stokes V-free configuration, we would expect to see a more edge-on disk 
signature in the H$\alpha$ emission line than that which is observed at the phase when the Stokes V signature is null.

Since a field with a surface dipolar strength lower than 200\,G would not confine the wind of HD\,155806 sufficiently to produce a disk; and since a stronger (but undetected) field would be in a highly improbable configuration that would in any case be unable to produce the emission structure seen in H$\alpha$, we conclude that it is unlikely that the disk of HD\,155806 is caused by a large-scale magnetic field confining the stellar wind.

\end{document}